\documentclass[preprint,12pt]{elsarticle}

\usepackage{color}
\biboptions{super,sort&compress}
\usepackage{parskip}
\usepackage{siunitx}

\usepackage{amssymb}

\usepackage{lineno}
\usepackage[hyphens]{url}

\usepackage{mdframed}
\usepackage[most]{tcolorbox}

\journal{arXiv}

\graphicspath{{Fig1/}{Fig2/}{Fig3/tps1/}{Fig4/}{Fig4/tps1/}}

\makeatletter 
\renewcommand\@biblabel[1]{#1} 
\makeatother

\begin{document}

\Urlmuskip=0mu plus 1mu

\begin{frontmatter}




\title{HIV transmission in men who have sex with men in England: on track for elimination by 2030?}


\author[label1]{Francesco Brizzi\fnref{fn1}}
\author[label1,label2]{Paul J Birrell\fnref{fn1}}
\author[label2]{Peter Kirwan}
\author[label2]{Dana Ogaz}
\author[label2]{Alison E Brown}
\author[label2]{Valerie C Delpech}
\author[label2]{O Noel Gill}
\author[label1,label2]{Daniela De Angelis\corref{cor1}}
\address[label1]{MRC Biostatistics Unit, University of Cambridge, Cambridge CB2 0SR, UK}
\address[label2]{National Infection Service, Public Health England, Colindale, NW9 5EQ, UK}
\cortext[cor1]{Correspondance to daniela.deangelis@mrc-bsu.cam.ac.uk}
\fntext[fn1]{Joint first authors}

\begin{abstract}

{\bf Background} After a decade of a treatment as prevention (TasP) strategy based on progressive HIV testing scale-up and earlier treatment, a reduction in the estimated number of new infections in men-who-have-sex-with-men (MSM) in England had yet to be identified by 2010. To achieve internationally agreed targets for HIV control and elimination, test-and-treat prevention efforts have been dramatically intensified over the period 2010-2015, and, from 2016, further strengthened by pre-exposure prophylaxis (PrEP).

\medskip

\noindent{\bf Methods} Application of a novel age-stratified back-calculation approach to data on new HIV diagnoses and CD4 count-at-diagnosis, enabled age-specific estimation of HIV incidence, undiagnosed infections and mean time-to-diagnosis across both the 2010--2015 and 2016--2018 periods.
Estimated incidence trends were then extrapolated, to quantify the likelihood of achieving HIV elimination by 2030.

\medskip

\noindent{\bf Findings} A fall in HIV incidence in MSM is estimated to have started in 2012/3, eighteen months before the observed fall in new diagnoses. A steep decrease from 2,770 annual infections (95\% credible interval 2.490--3,040) in 2013 to 1,740 (1,500--2,010) in 2015 is estimated, followed by steady decline from 2016, reaching 854 (441--1,540) infections in 2018. A decline is consistently estimated in all age groups, with a fall particularly marked in the 24-35 age group, and slowest in the 45+ group.
Comparable declines are estimated in the number of undiagnosed infections. 
\medskip

\noindent{\bf Interpretation} The peak and subsequent sharp decline in HIV incidence occurred prior to the phase-in of PrEP. Definining elimination as a public health threat to be $<50$ new infections (1.1 infections per 10,000 at risk), 40\% of incidence projections hit this threshold by 2030. In practice, targeted policies will be required, particularly among the 45+y where STIs are increasing most rapidly. 

\medskip
\noindent{\bf Funding} Medical Research Council; UK National Institute of Health Research Health Protection Units on Evaluation of Interventions; Public Health England.

\end{abstract}


\end{frontmatter}

\section*{Background}
\label{sec:intro}

\noindent In 2015 Member States of the United Nations adopted the Sustainable Development Goals, including the target to end the AIDS epidemic by 2030.\cite{UNDP15} This followed the 90-90-90 targets (90\% of people living with HIV being diagnosed; 90\% of diagnosed individuals receiving anti-retroviral therapy (ART); and 90\% of people on ART being virally suppressed by 2020)\cite{WHO16} set in 2014 by the Joint United Nations Programme on HIV/AIDS (UNAIDS) as a fast-track strategy towards HIV elimination.\cite{UNAIDS15} In England, regular monitoring of HIV prevalence and care has shown the 90-90-90 target to have been achieved among men-who-have-sex-with-men (MSM) by the end of 2016,\cite{PHE18} with a sustained decline in the number of new HIV diagnoses observed in this group since 2015.\cite{PHE18} The goal now set for the country is to reach HIV elimination by 2030.\cite{UKGOV19}

Monitoring HIV elimination in MSM requires regular assessments of the incidence of HIV infection. Although trends in new HIV diagnoses provide a valuable metric, new diagnoses in any given time period do not equate to the number of new infections in that period. Diagnoses are a dynamic mixture of long-standing and recent HIV infections, resulting from the interplay between transmission, infection progression and diagnosis. Only by modelling these unobserved processes is it possible to disentangle the contributions of changes in testing intensity and transmission to the observed HIV diagnoses and thus correctly estimate the underlying number of new infections. The CD4-staged back-calculation approach achieves this by using information on new HIV and AIDS diagnoses, CD4 counts around diagnosis and the natural history of HIV infection to reconstruct HIV incidence and uncover time-varying diagnosis rates.\cite{SweDA05} In addition, estimates are provided of the number of undiagnosed infections over time and trends in the time interval from infection to diagnosis.\cite{BirCGDD12}

Over the 2001--2010 period, despite a large and steady increase in HIV testing, greater retention in care and improvements in the proportion of people living with HIV who received ART, a back-calculation analysis showed persistent high HIV incidence among MSM in England of around 2,500 new infections annually.\cite{BirGDBDCRD13} This finding was consistent with independent studies estimating increasing incidence in MSM over the periods 2000–2010 and 2000–2014 periods.\cite{PhiCNBLRMEHJLD13,NakVTSRCAABCCDFKLLPPQTTDP16} It is likely that the HIV epidemic was contained somewhat by testing and treatment during this time of resurgence in condom-less sex and wider partnership formation facilitated by increasing online connectivity.\cite{AghWNPCMHGJ16,LogFBRHMN18,SavMDIZH12} Nevertheless, the plateau in HIV incidence was described as “sobering”, giving rise to doubts that testing-and-treatment efforts could substantially reduce HIV transmission in MSM.\cite{Gra13}

Since 2010, there have been important changes to the biomedical components of combination HIV prevention for MSM in England.\cite{UNAIDS10} Initially, treatment as prevention (TasP) had been advocated following mathematical modelling that showed the impact of universal test-and-treat strategies on HIV incidence;\cite{GraGDDW09} ecological studies that supported an inverse relationship between ART coverage and the number of new diagnoses;\cite{MonLBYWKSHHDK10} and evidence from the HPTN 052 trial that early ART initiation reduced sexual transmission in sero-discordant heterosexual couples.\cite{CohCMGHKHKGPGMCSMHEPWMMBSEGHSREBTNCEF11} After 2010 HIV testing was intensified and treatment scaled up. Three-monthly (rather than six-monthly) HIV tests for most-at-risk MSM were recommended in 2012,\cite{CluFBWNHKFS12} and new guidelines were issued to offer ART to those with a CD4 count $<350$ cells mm$^{-3}$.\cite{BHIVA12} In 2015 immediate ART initiation was recommended for all persons newly diagnosed with HIV infection.\cite{BHIVA15,BroMOKYNFHCDG17}

In addition, from 2015, the numbers of MSM accessing pre-exposure prophylaxis (PrEP) began increasing. Following evidence on the efficacy of PrEP in MSM when taken consistently by high-risk individuals,\cite{GraLAMLGCGRMFVBCSBMKAMBHGDPWMZLRJMBG10} the PROUD trial was carried out in England,\cite{DolDMGCFSSMRPSFBBBLTWAGMGDN16} resulting in some high-risk MSM initiating PrEP in 2012. Around 25 were taking PrEP by the end of 2012, 250 by the end of 2013, and 500 by the end of 2014. In Autumn 2015, internet sites were established to facilitate self-purchase of PrEP from abroad. An online survey of MSM indicated that PrEP usage quadrupled during 2016 so that an estimated 3,000 MSM were taking PrEP by year end.\cite{PrEP18} Following the start of a large-scale PrEP implementation trial in the Autumn of 2017, estimated numbers on PrEP progressively increased to around 5,000 by end 2017 and 15,000 by end 2018.\cite{NHS19}

Consensus has yet to be reached on a definition of HIV elimination in the MSM community in England. A possible elimination threshold could be fewer than 50 new infections acquired annually within the country. Regardless of how it is defined, the prospect of achieving elimination will depend on the lessons to be learnt about the impact of the steady amplification of combination prevention over the past ten years. These lessons might be different in different subgroups of the MSM population. Crucially, HIV progression depends on age at infection,\cite{CAS00a} time to treatment varies substantially between age groups,\cite{PHE16} and trends in the number of new diagnoses and bacterial sexually transmitted infections (STIs) observed since 2010 also differ by age.\cite{BroNCKOCDD18} This heterogeneity could be reflected in a differential impact of combination prevention across age groups.


In this paper we reconstruct trends in both overall and age-specific HIV incidence over the period 2010-2018 for MSM in England through the application of a novel back-calculation model to the latest diagnosis data.\cite{BriBPKBDGD19} We investigate whether these trends mirror changes in the biomedical components of combination prevention during pre-PrEP (2010-2015) and post-PrEP (2016-2018) periods, estimate the timing of any peak in infection, and examine whether incidence changes were consistent across age groups. In addition, assuming current trends are maintained in the near future, we venture to predict incidence up to 2030, assessing whether elimination by this data is a realistic goal.

\section*{Data and Methods}

\subsection*{Data Sources}

To provide a comprehensive description of the epidemic, we extracted data on HIV testing among MSM from the GUMCAD STI surveillance system,\cite{SavMDIZH12} alongside data on the number of diagnoses of bacterial STIs such as gonorrhoea and syphilis.

Information on quarterly numbers of new HIV and AIDS diagnoses was obtained, together with the age at diagnosis, linked data on CD4 counts taken within three months of an initial diagnosis, and the time of initiation of ART, from the HIV and AIDS Reporting System (HARS). Here an AIDS diagnosis is an HIV diagnosis followed by AIDS defining symptoms within three months of the initial diagnosis.


\subsection*{Back-calculation Analyses}

The quarterly HIV and AIDS diagnosis and CD4 data were analysed using a novel model that extends the CD4-staged back-calculation previously used to monitor the HIV epidemic.\cite{SweDA05, BirCGDD12} More specifically, this new approach allows the estimation of age- and time-specific incidence (using the thin-plate regression smoothing spline of Wood\cite{Woo03}) and the stratification of the infected and undiagnosed MSM population into compartments defined by CD4 count, current age and age at seroconversion.\cite{BriBPKBDGD19} Movement out of these compartments is either due to infection progression (i.e. move to a lower CD4 compartment) or to diagnosis (move outside of the undiagnosed population). Model outputs, obtained using Bayesian inference, include posterior distributions for: quarterly age-stratified numbers of new infections, diagnosis probabilities, and numbers of undiagnosed infections. Results are informed by the observed HIV, AIDS and CD4 data, age (at infection) specific infection progression probabilities estimated from cohort studies, and prior distributions on model parameters. See Brizzi et al. for further details.\cite{BriBPKBDGD19}

\subsection*{Incidence Estimation and Projection}
Estimates of incidence are obtained by dividing the estimated number of new infections by the size of the uninfected MSM population, derived through recent work on HIV prevalence in England.\cite{PreKMCHHJMBDGD19} To predict incidence beyond 2018, the estimated age-time profile for the number of new infections resulting from the smoothing spline is projected forward under the assumption that current incidence trends persist. The denominator population size is assumed to remain constant from 2018 onwards.


\subsection*{Role of the funding source}
None of the bodies providing funding exerted any influence over the design of the study. The corresponding author had access to all data used in the analyses presented here and made the final decision to submit the manuscript for publication.

\section*{Results}

\begin{figure}[ht]
\centering
{\bf \small A}\hspace{0.5\linewidth}{\bf \small D}\\
\includegraphics[height=0.2489242\textheight]{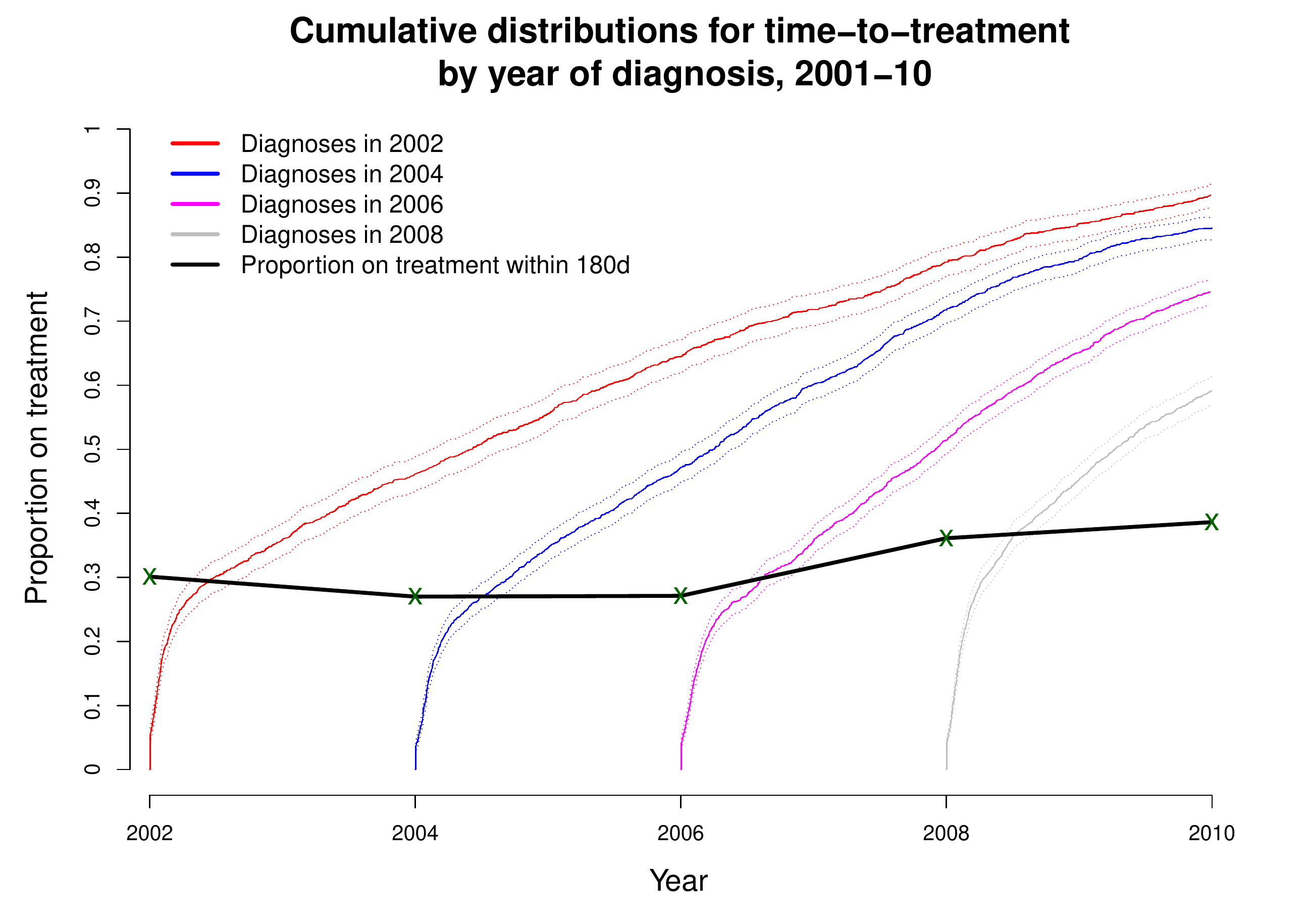}
\includegraphics[height=0.2489242\textheight]{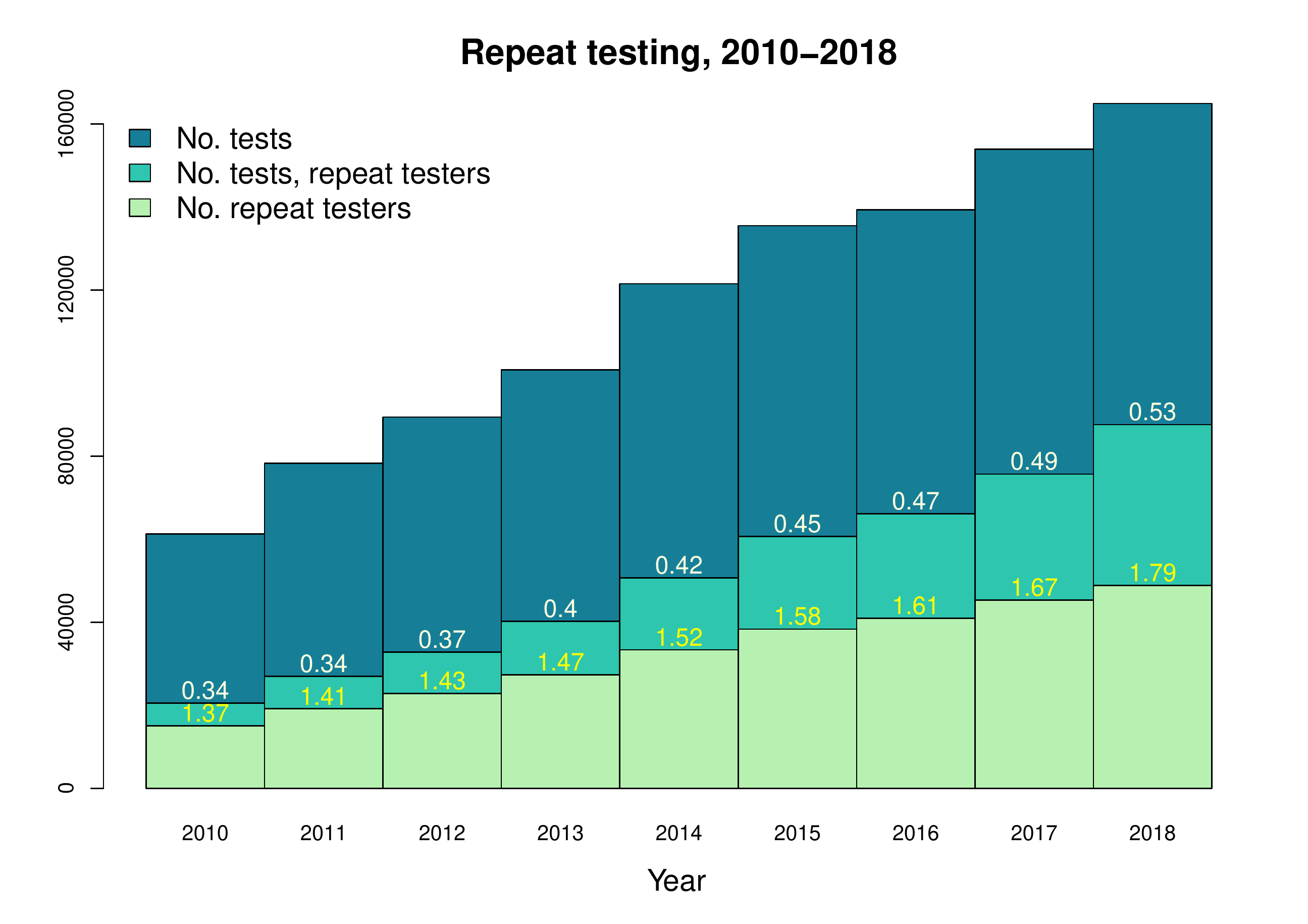}\\
{\bf \small B}\hspace{0.5\linewidth}{\bf \small E}\\
\includegraphics[height=0.2503125\textheight]{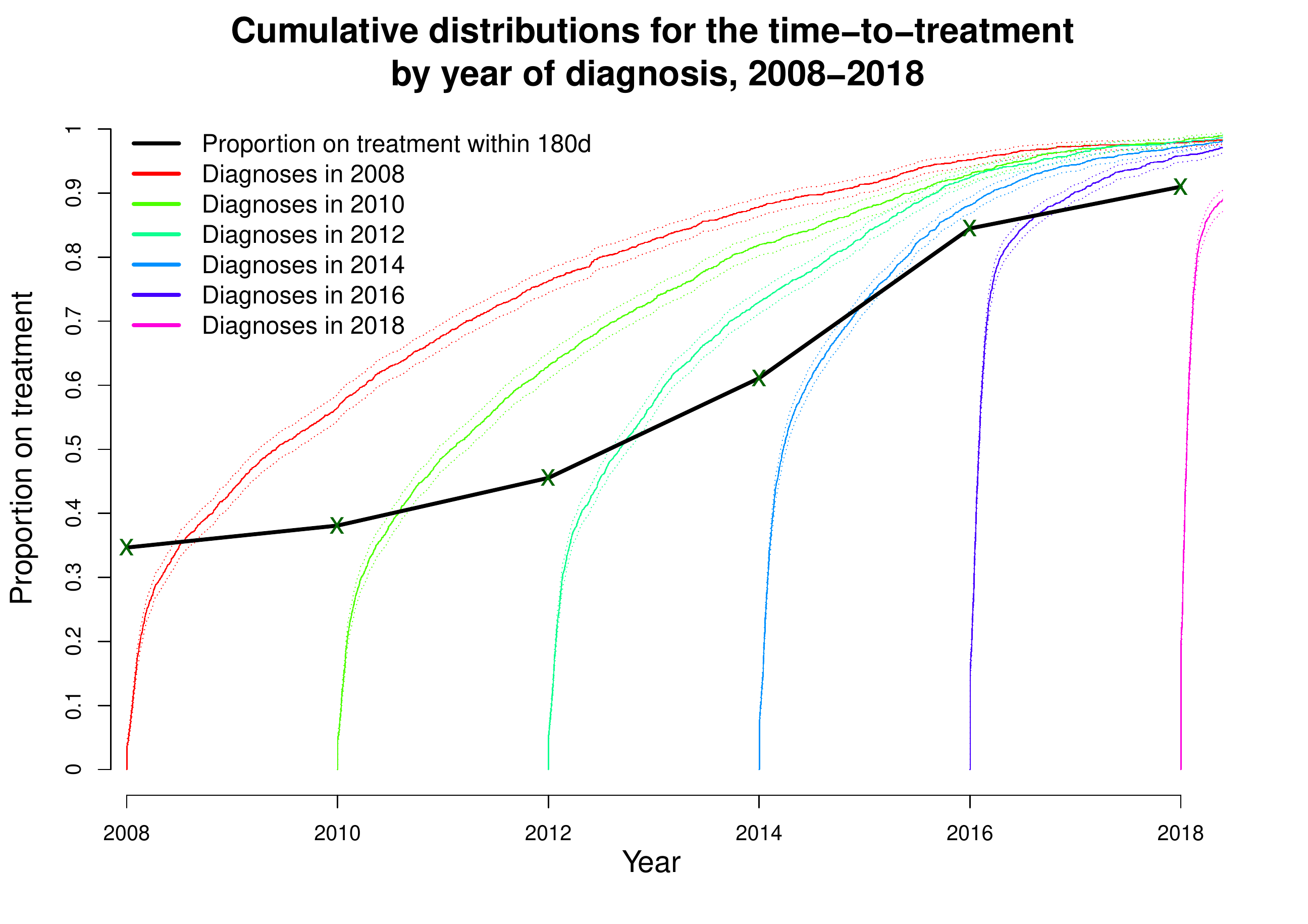}\includegraphics[height=0.2503125\textheight]{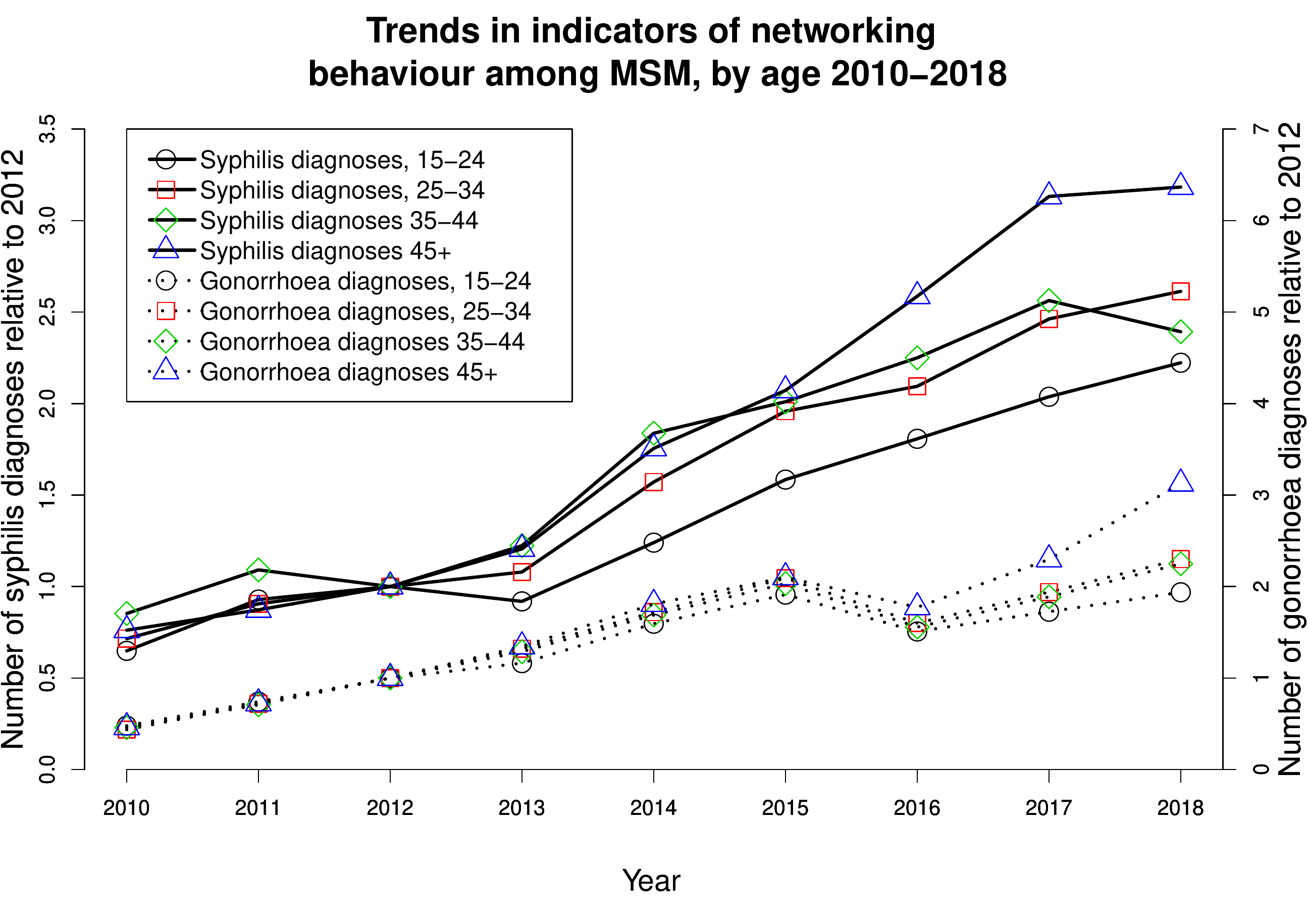}\\
{\bf \small C}\hspace{0.5\linewidth}{\bf \small F}\\
\includegraphics[height=0.2503125\textheight]{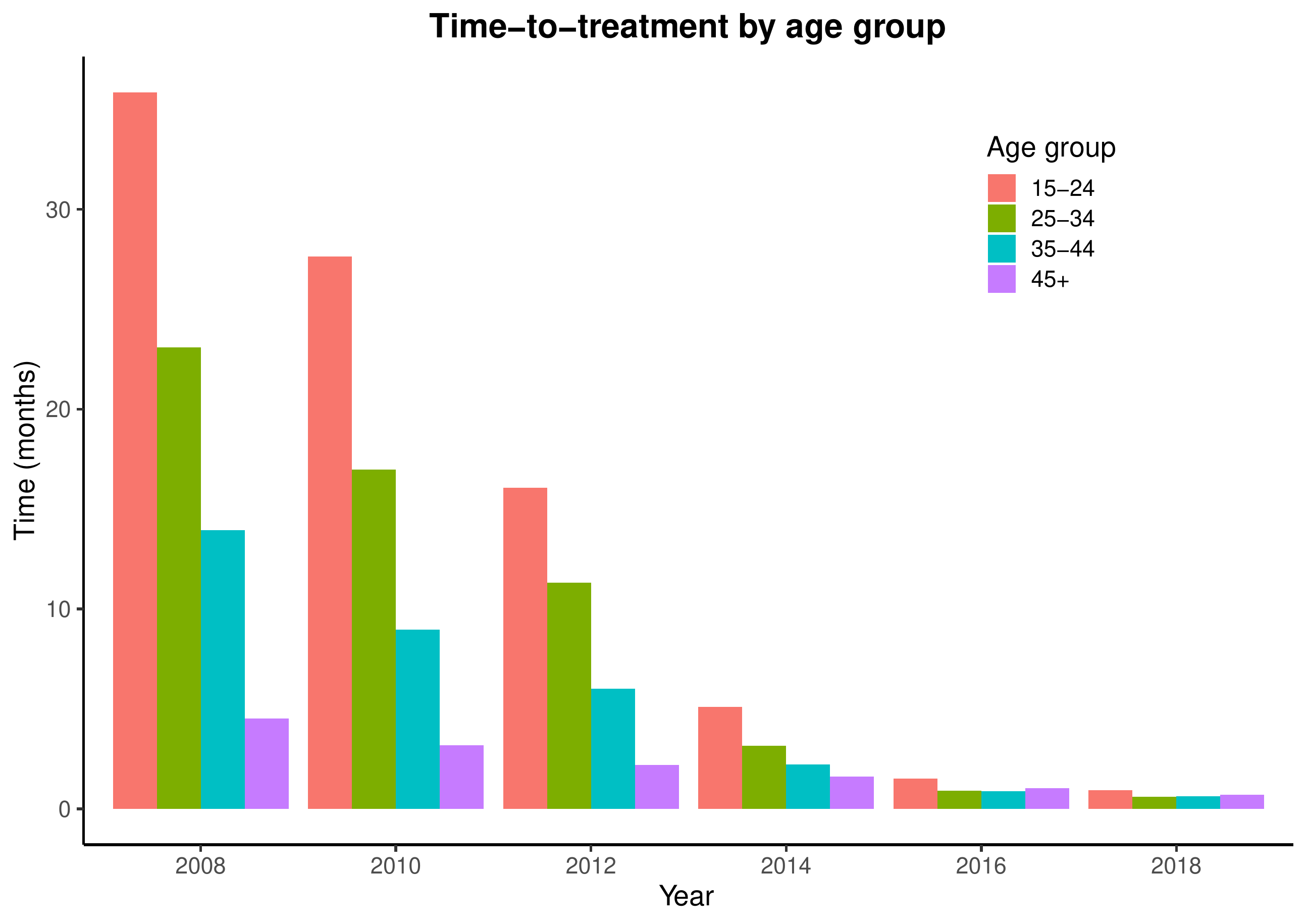}\includegraphics[height=0.2503125\textheight]{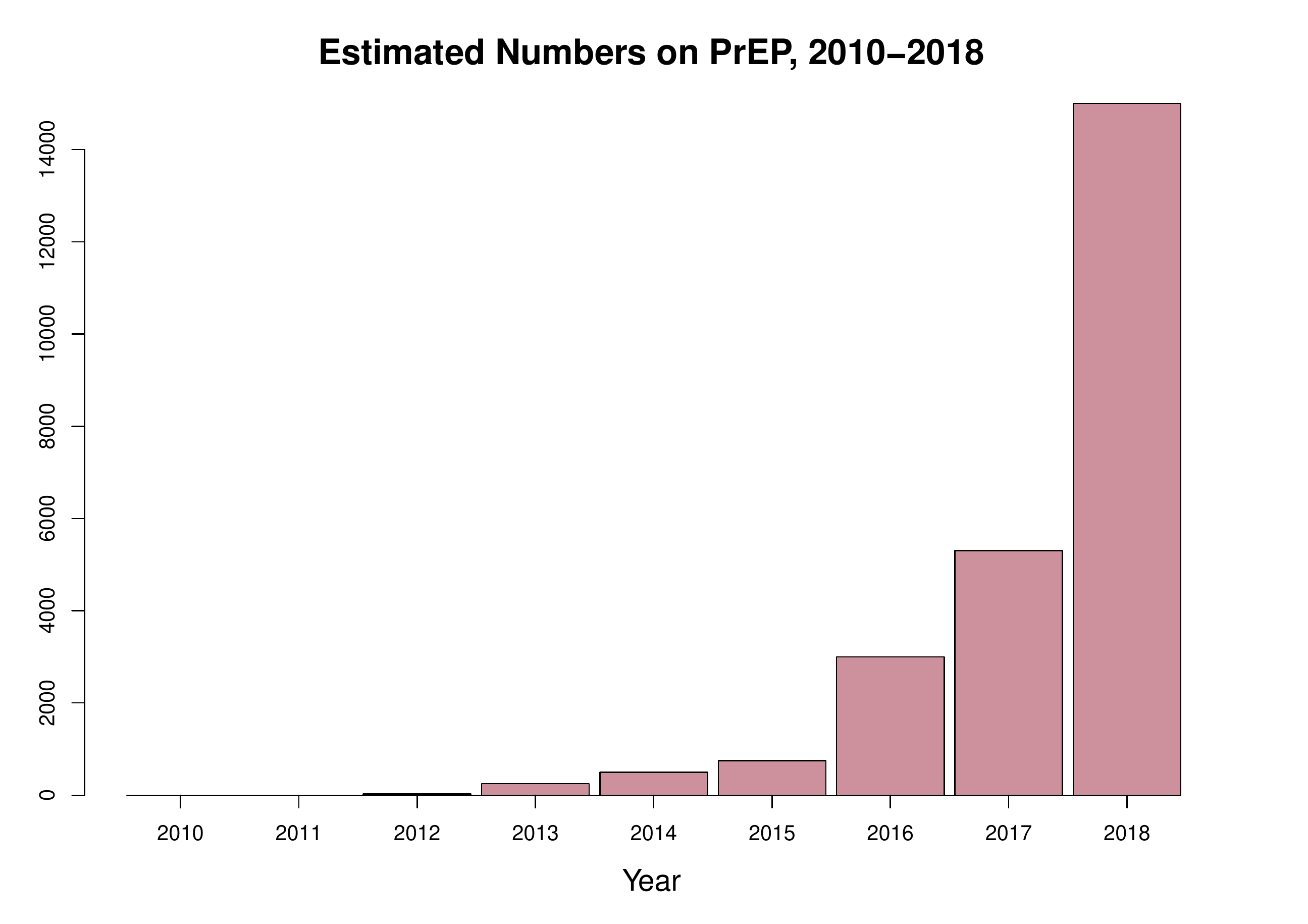}
\caption{\scriptsize (A) Cumulative distributions for the time to ART initiation for MSM diagnosed with HIV in 2002, 2004, 2006 and 2008. 
The black line gives the proportion of MSM diagnosed with HIV on ART within 180 days of diagnosis;
(B) Annual numbers of HIV tests in MSM in STI clinics, the most common setting for HIV tests in MSM. Numbers on top of the light green bars give the average number of tests taken in a year by a repeat tester. Numbers on top of the dark green bars give the fraction of testers classified as `repeat’ testers.
(C) As Panel A, for years 2008, 2010, 2012, 2014, 2016 and 2018.  (D) Annual number of syphilis (left-hand axis) and gonorrhoea diagnoses in MSM (right-hand axis) by age group relative to the 2012 number of diagnoses in that age group;
(E) Mean time to ART initiation following diagnosis by age group in the years 2008--2018; (F) Scale-up in the number of people on PrEP since 2010. 
}
\label{fig:new.era}
\end{figure}

The left column of Fig. \ref{fig:new.era} presents changes in the distribution of time-to-treatment initiation over the periods from 2002 through 2010 (Panel A) and from 2010 to 2018 (Panel B). Evidently much greater improvement was achieved beyond 2010, with the cumulative distributions of the time-to-treatment being close to vertical for diagnoses in 2016 and 2018, indicating near immediate initiation of ART. The black line in both plots indicates the percentage of MSM on ART within 180 days of diagnosis rising from \num{34.7}\%
in 2008 to \num{84.5}\% in 2016 and \num{91.0}\% in 2018. Panel E provides the change in the median time-to-treatment by age, showing that the observed improvement is mostly attributable to faster access to treatment in the 15-24 (median time of \num{35.9} months in 2008 decreasing to \num{0.92} months in 2018) and 25-34 (\num{23.1} decreasing to \num{0.59} months) age groups. The over 45 age group already had a short waiting time-to-treatment (\num{4.53} decreasing to \num{0.70} months) and did not have the same capacity for significant improvement.

Patterns of HIV testing and new diagnoses following the testing and treatment scale-up after 2010 are presented in Fig. \ref{fig:new.era} Panel B, showing year-on-year increases in the annual number of tests, the proportion of tests taken up by repeat testers (MSM with at least two tests within a 365-day period) and the overall number of repeat testers.  This greater testing effort is set against a steady rise in the incidence of bacterial STIs from 2010 to 2018 (Panel D), with gonorrhoea and syphilis diagnoses increasing at a rate of over 2,000 and 500 diagnoses per year, respectively. Note that the plot in Panel D has been scaled to show the relative increases in diagnosis of the two STIs across age-groups. Both infections show a higher increase in the over 45 age group. Panel F illustrates the scale of the PrEP phase-in quantified in the Introduction.

\begin{figure}
    \centering
    {\bf A}\\
    \includegraphics[width=.8\linewidth]{\detokenize{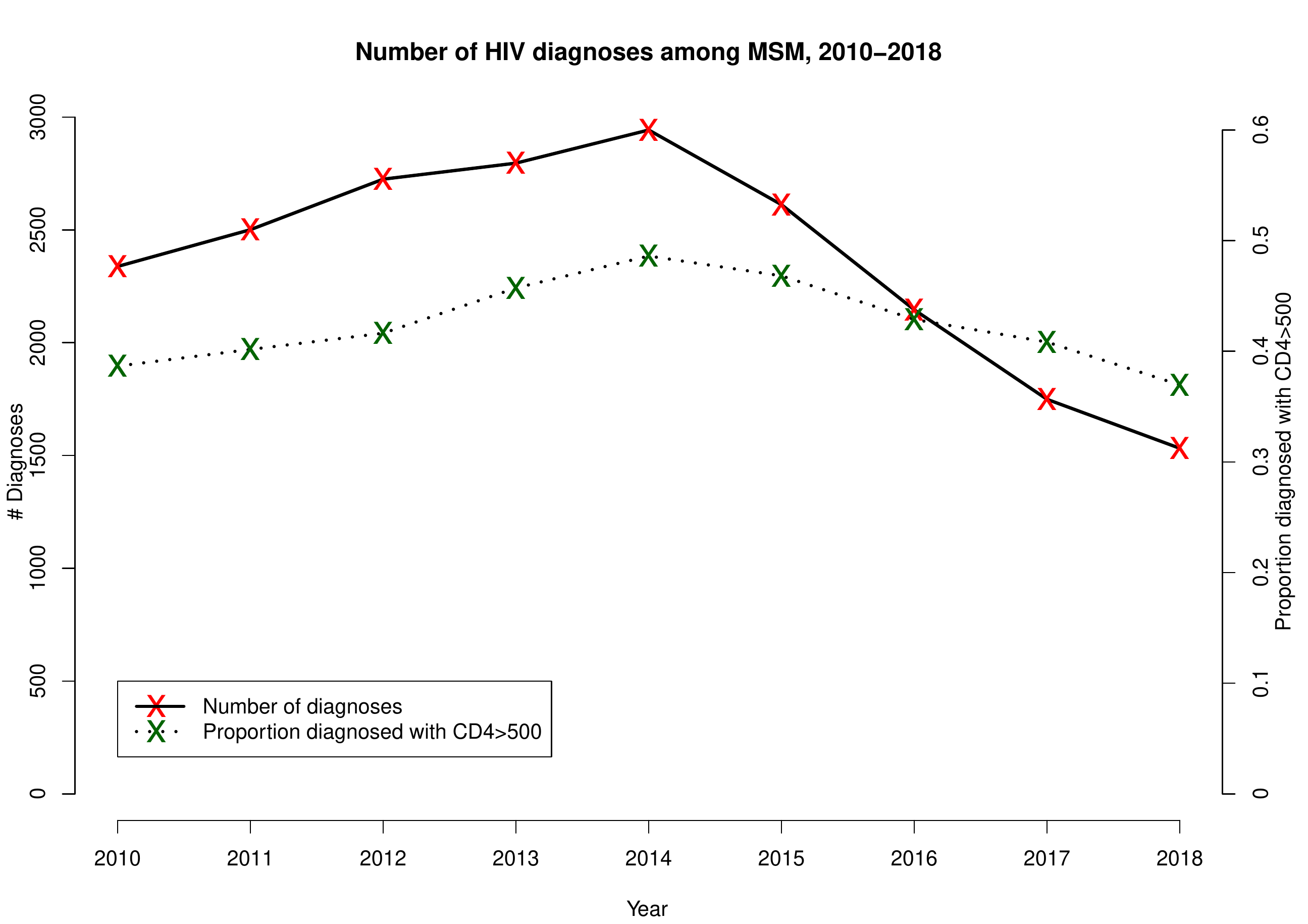}}\\
    {\bf B}\\
    \includegraphics[width=.8\linewidth]{\detokenize{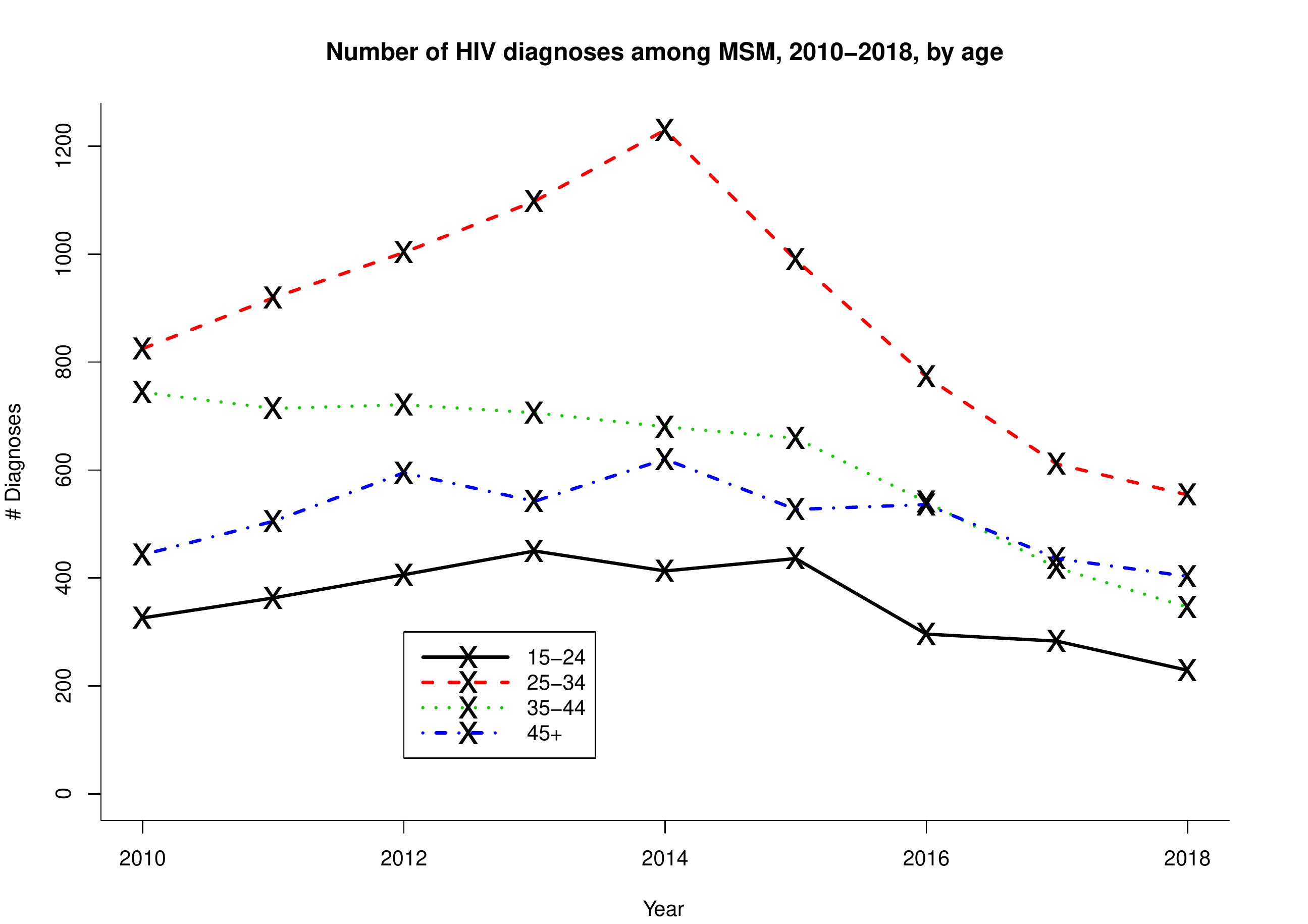}}
    \caption{(A) Total annual number of HIV Diagnoses, and proportion of HIV diagnoses with accompanying CD4 blood cell count within three months; (B) Annual number of HIV diagnoses by age-groups 15--24, 25--34, 35--44 and 45+.}\label{fig:data}
    \label{fig:my_label}
\end{figure}

In contrast to the increase in STI diagnoses during 2010--2018 are the trends in the number of new HIV diagnoses (Fig. \ref{fig:data}, Panel A) which rise to a peak in 2014 before displaying a sharp decline. Stratification by age group (Panel B) reveals a modest though sustained decline over the whole period in the 35–44 age group, contrasting with the sharp downturn in 2014 amongst 25–34-year-olds. At this time, more modest declines also begin in the 15-24-year-olds and eventually in the 45+ age group.

\begin{figure}[ht]
\centering
{\bf A}\hspace{0.48\linewidth}{\bf C}\\
\includegraphics[width=0.495\linewidth]{\detokenize{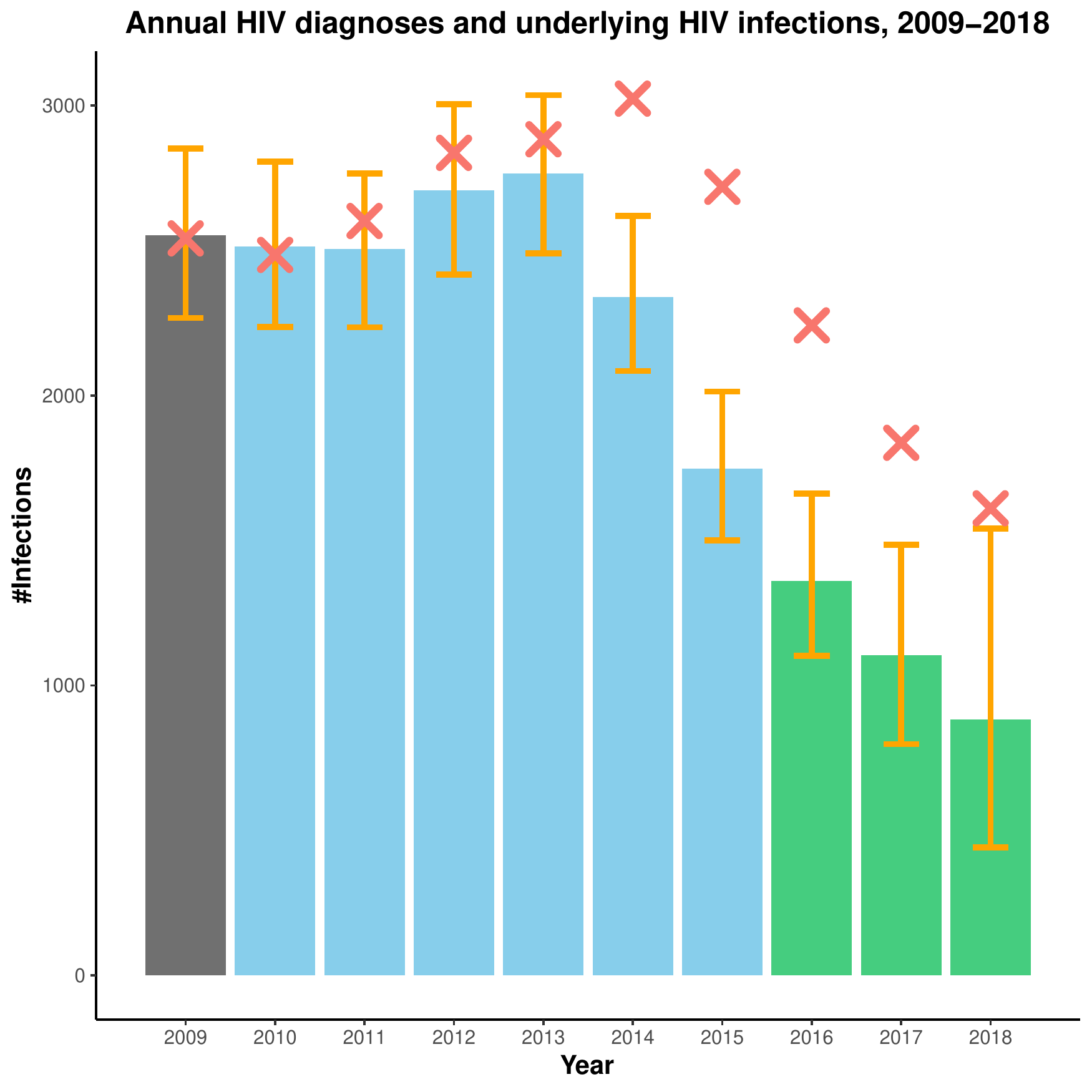}}\includegraphics[width=0.495\linewidth]{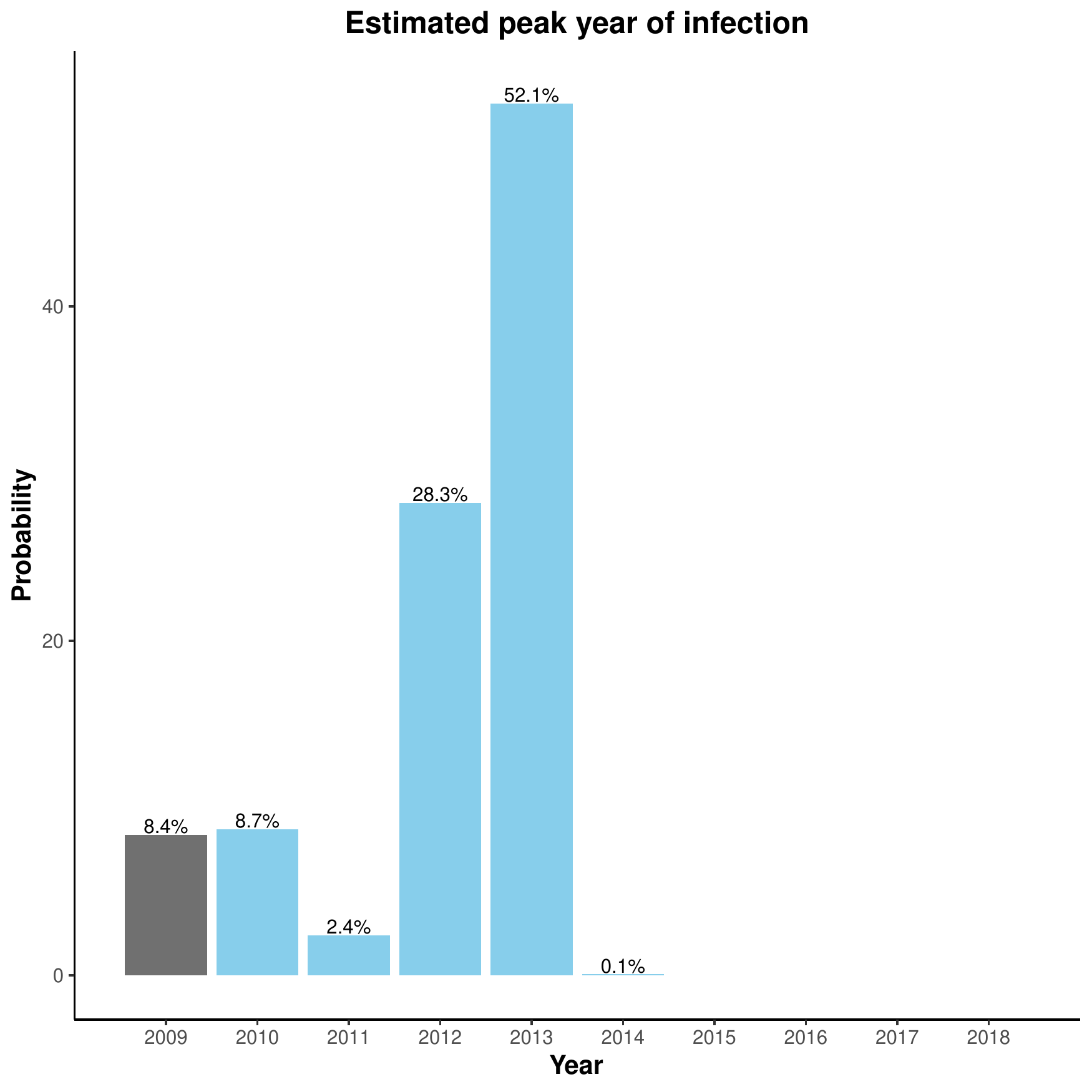}
{\bf B}\hspace{0.48\linewidth}{\bf D}\\
\includegraphics[width=0.495\linewidth]{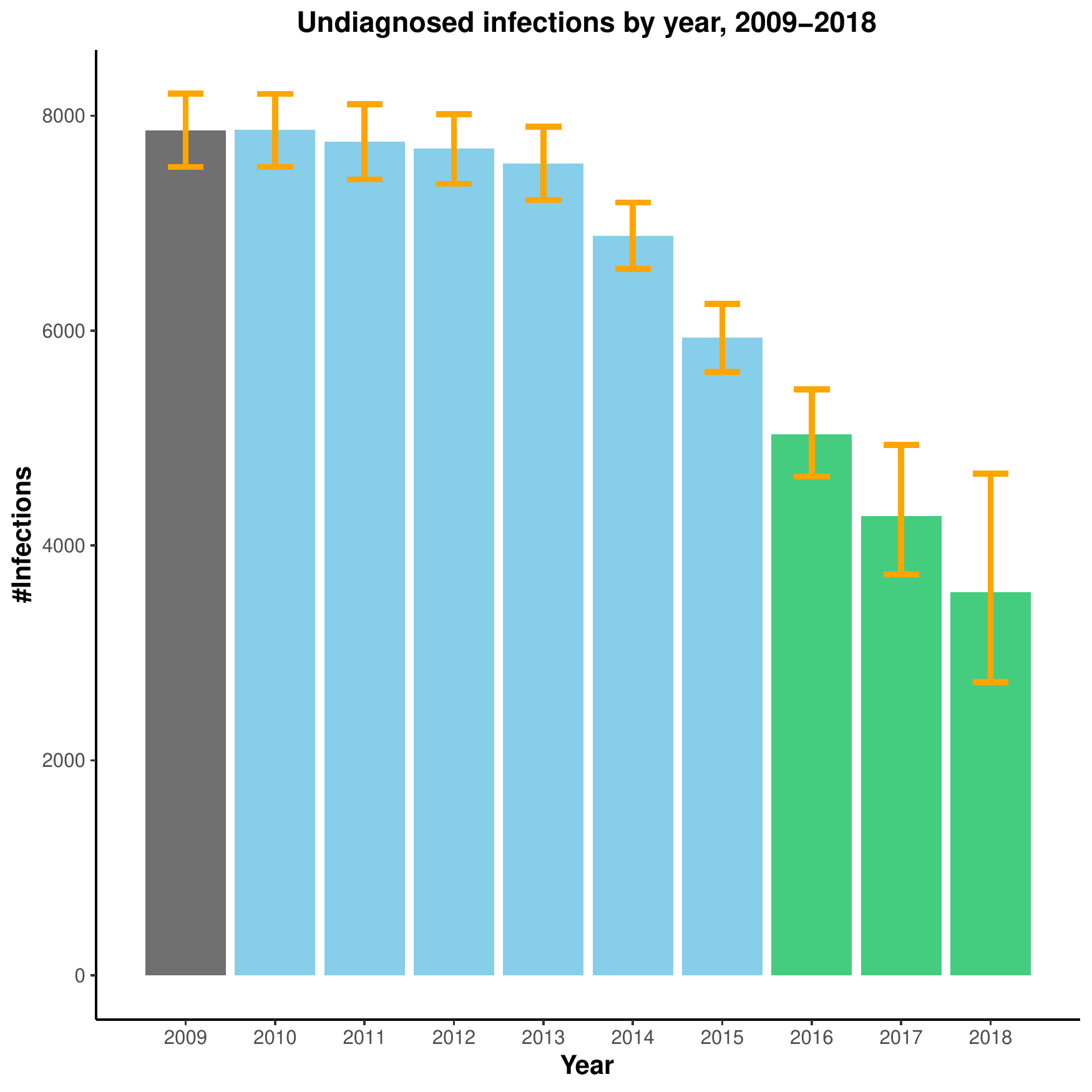}
\includegraphics[width=0.495\linewidth]{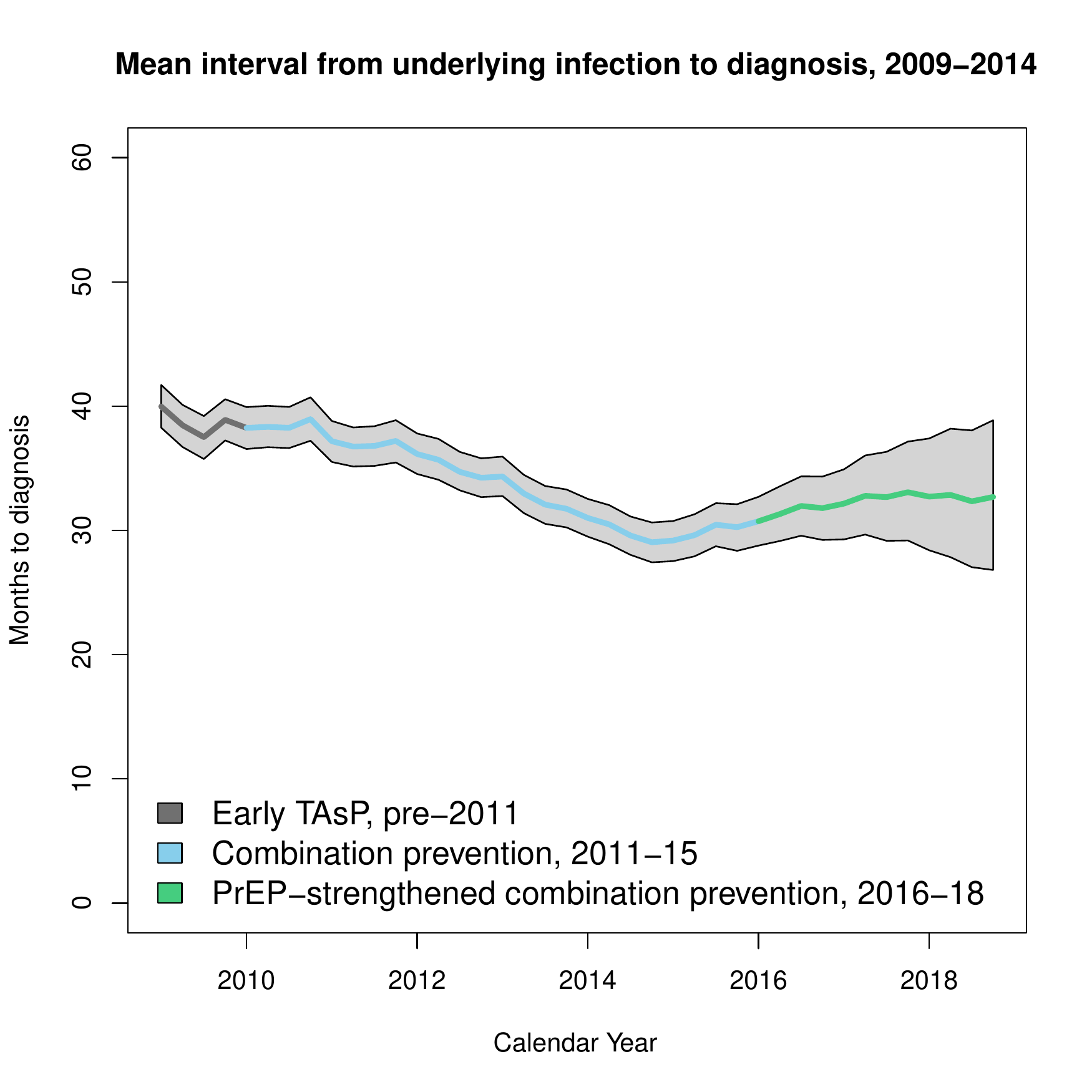}\\
\caption{(A) Estimated total annual number of new HIV infections, with error bars denoting 95\% credible intervals. The crosses represent the observed number of diagnoses in each year; (B) Distribution of the estimated peak year in the number of new HIV infections; (C) Estimated number of undiagnosed infections in MSM in England by year (with 95\% credible intervals); (D) The plotted line gives the estimated mean time interval between HIV infection and diagnosis, measured in months. The shaded area gives 95\% credible intervals for the mean.}\label{fig:ai.estimates}
\end{figure}

Fig. \ref{fig:ai.estimates} summarises overall estimates of new (Panel A) and prevalent undiagnosed infections (Panel B) corresponding to eras of intensified combination prevention (2010--2015, light blue), and the PrEP-strengthened combination prevention (2016--2018, green). Panel A shows a steep decline in the annual number of new infections from 2,770 (95\% credible interval, 2,490--3,040) in 2013 to 1,740 (1,500--2,010) in 2015. In the PrEP-strengthened era the decline continues, with 854 (441--1,540) new infections in 2018. The timing of the peak in infections (Panel B) is estimated with 80\% certainty to be in either 2012 or 2013, with 2013 being the most likely peak year. The estimated number of MSM living with undiagnosed HIV infection shows a gradual decline beginning at the start of the intensified combination prevention era, with 7,880 (7,540--8,220) undiagnosed infections. The decline accelerates from 2013, falling sharply from 7,700 (7,490--8,040) infections at the beginning of 2013 to 5,930 (5,370--6,030) infections at the end of 2015, and then down to a low of 3,530 (2,730--4,670) undiagnosed infections at the end of 2018, a fall of 40.5\% since end 2015.

Fig. \ref{fig:ai.estimates}, Panel D shows an estimated mean time between infection and diagnosis of \num{40.0} (\num{38.3}--\num{41.7}) months in 2009, declining to a low of \num{29.0} (\num{27.4}--\num{30.6}) months at the end of 2014 before increasing slightly to \num{32.7} (\num{26.8}--\num{38.9}) months by the end of 2018. This apparent recent increase is subject to substantial uncertainty and is linked to the decreasing proportion of infections diagnosed at CD4 cell count $>$500 cells mm$^{-3}$.

\begin{figure}[ht]
\centering
{\bf A}\\
\includegraphics[height=0.425\textheight]{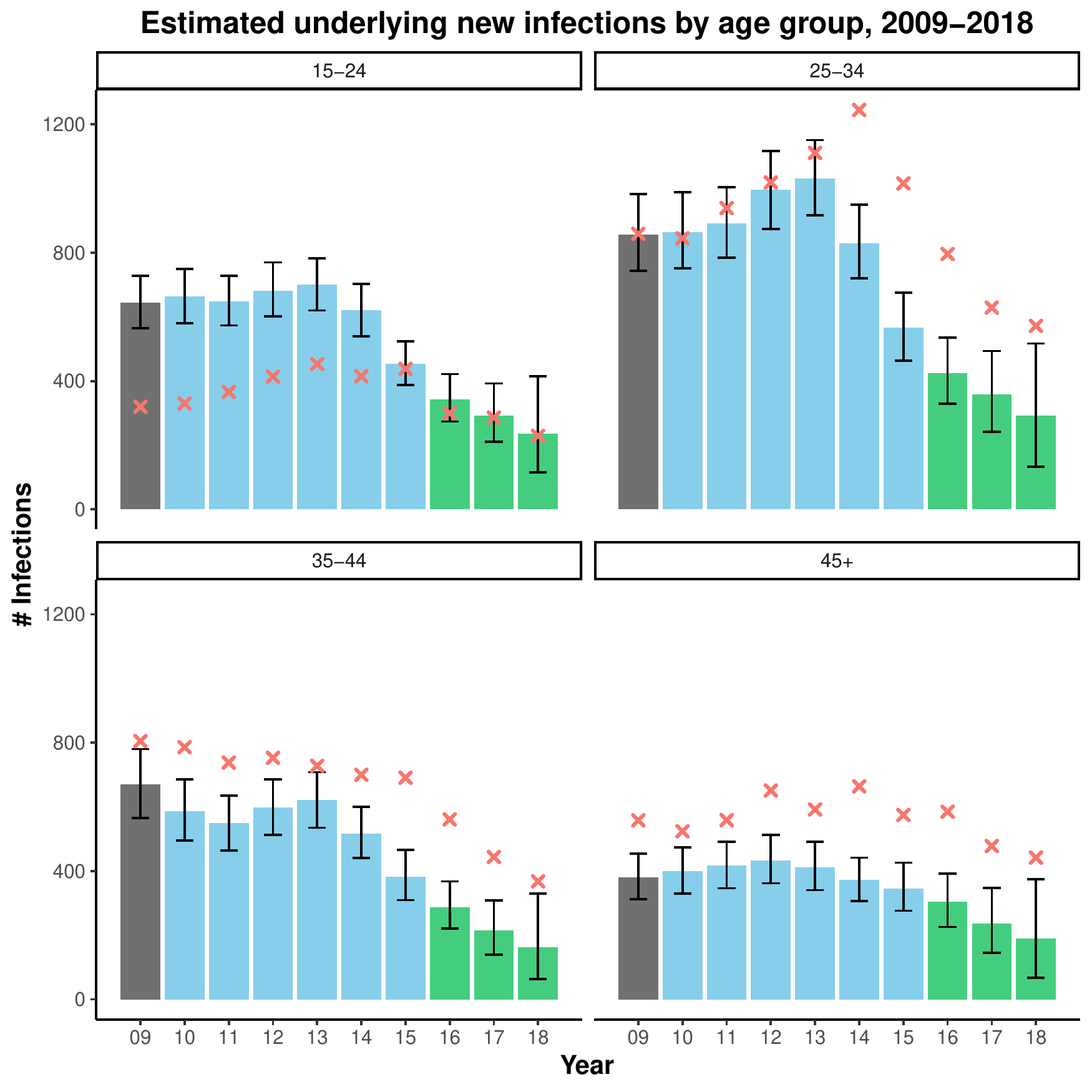}\\
{\bf B}\\
\includegraphics[height=0.425\textheight]{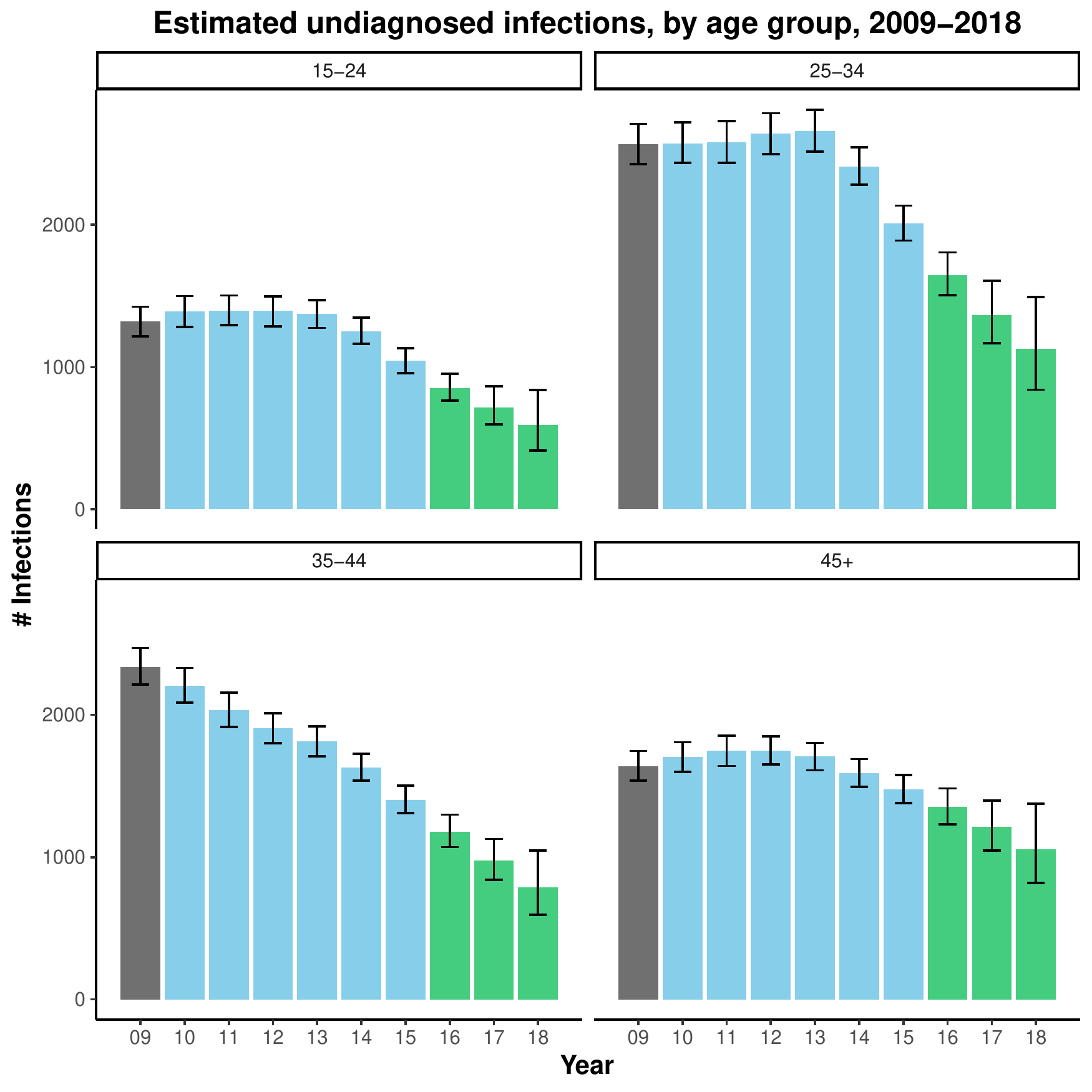}
\caption{Estimated annual number of new HIV infections (Panel A) and undiagnosed infections (Panel B), with associated 95\% credible intervals in each of the four age groups. Crosses in (A) again represent the observed diagnosis data in each year.}\label{fig:ad.estimates}
\end{figure}

When examined by age group, the estimated number of new HIV infections follows different trends. Infections in the 15--34 year-old MSM (Fig. \ref{fig:ad.estimates}, Panel A) show a pattern consistent with what is observed overall (Fig. \ref{fig:ai.estimates}, Panel A), with a steep decline from 2013 onwards; in the 35--44 age group the number of new infections oscillates prior to 2013 when a similar decline begins; and in the 45+ the decline is much more gradual.  Age heterogeneity is also observed in the estimated number of undiagnosed infections with the 15--34 age group displaying a decline similar to the new infections in this group; undiagnosed infections in the 35--44 year-old group steadily declining; and trends in the 45+ being again more gradual, so that in this group in 2018 the number of infected undiagnosed MSM is higher than in the 35--44 group.

\section*{Discussion}


This reconstruction of HIV incidence in the MSM population in England during the period 2010--2018, shows the epidemic to be waning rapidly despite large increases in the incidence of bacterial STIs. During the period 2011--2015 the continuation of intensification of HIV testing together with the rapid adoption of immediate treatment following diagnosis, corresponded to a downturn in the number of new HIV infections with a decline of 37\% from 2012 to 2015.  Since the expansion of PrEP access that began in late 2016 there have been further annual decreases in the number of new infections of around 20\%. 

The peak in the number of infections occurred over the years 2012--2013, preceding the peak in HIV diagnoses by at least eighteen months. Importantly, the downturn in infection begins well ahead of the start of PrEP scale-up and the 20\% year-on-year decrease in infections since 2016 is at a similar rate of decline as for the previous two years. The additional impact of the major expansion of PrEP through 2018, though likely a contributor to this decline, is difficult to discern, set against the continued intensification of testing and immediate treatment. Nevertheless, the extra impetus of the introduction of a national PrEP programme in 2020 may cause an acceleration in incidence decline to become apparent,\cite{NHS16} or at least ensure the decline is maintained at its recent rate. 

The fall in incidence is consistent in all age groups, being particularly marked in the 24-35 age group, and slowest in the 45+ age group. It has been indicated that the 45+ stood to benefit the least from the policy of immediate ART treatment for new HIV diagnoses, whilst this group reported the greatest proportional increase in new cases of syphilis and gonorrhoea. This under-served group could benefit through focussed efforts to increase the number of regular HIV testers.

An estimate of the HIV incidence rate overall can be derived by using estimates of the number of uninfected MSM. Current work on trends in HIV prevalence in England provides, from 2012 onward, estimates for these denominators which steadily increase from around 456,000 in 2012 to 473,000 in 2018.\cite{PreKMCHHJMBDGD19} In terms of incidence rate, our results correspond to a decrease from \num{59.3} infections per 10,000 at risk (95\% CrI: \num{49.6}--\num{71.0}) to \num{37.5} (\num{30.4}--\num{46.2}) over the 2012--2015 period, with a further decrease to \num{18.0} (\num{9.20}--\num{33.0}) up to the end of 2018.

When compared with most high-income countries, this incidence fall is impressive. In both the USA and Australia, estimated HIV incidence in MSM has remained constant at 50 and 80 per 10,000 over the 2012--2016 and 2008--2015 periods respectively.\cite{SinSJMH18,Kirby17} and in Canada a small decline from 50 to 40 between 2005 and 2014 has been estimated with a stabilisation of incidence to the end of 2016.\cite{YanZ18} 

\begin{figure}
    \centering
    \includegraphics[width=\textwidth]{\detokenize{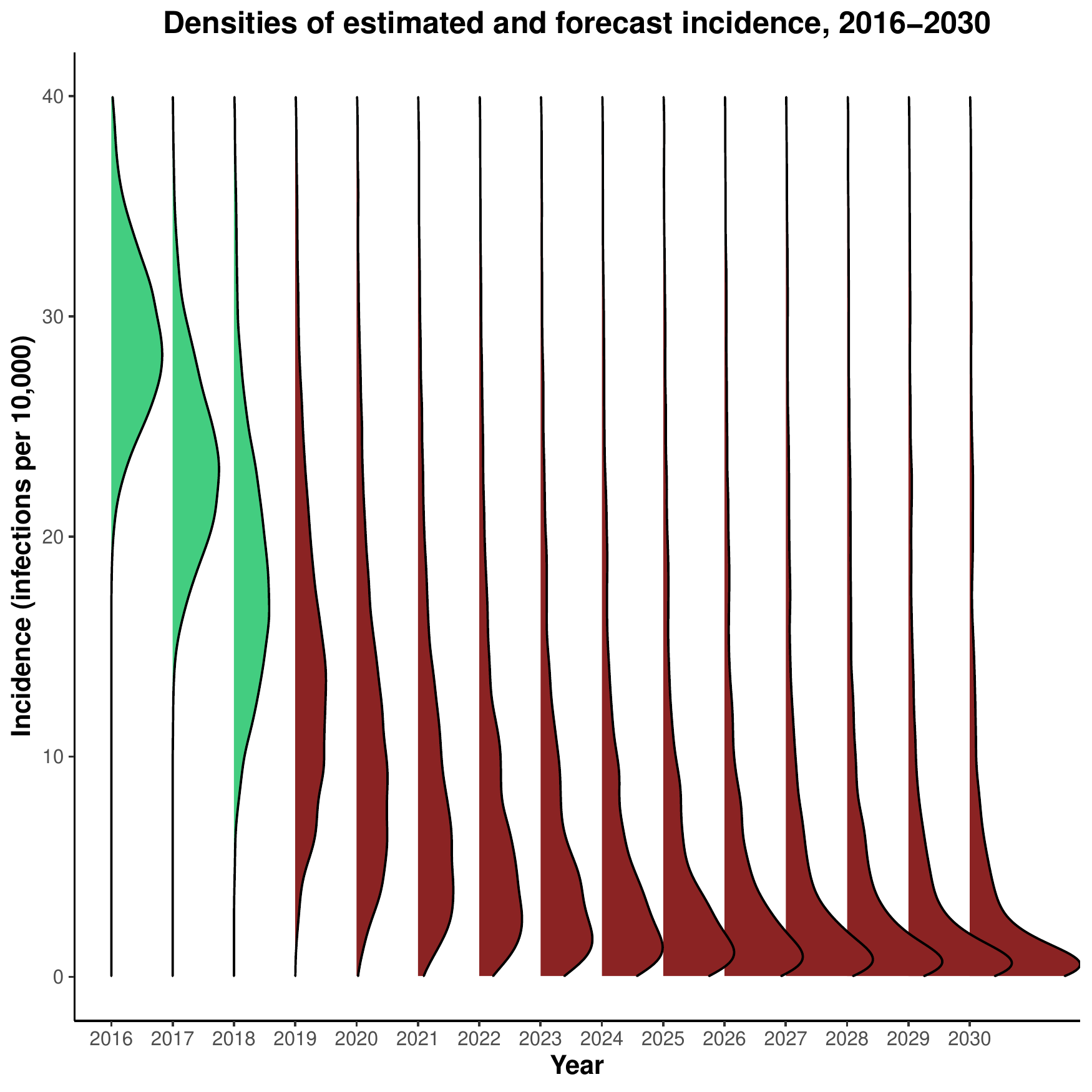}}
    \caption{Density plots of estimated incidence in years 2016--2018, with corresponding distributions for the projected incidence in each of the years over the period 2019--2030.}\label{fig:projections}
\end{figure}

Our results suggest elimination of HIV as a public health threat for MSM in England in the near future may be within reach. To assess this prospect, we extrapolated the recent trends in new infections and predict a greater-than-ten-fold fall from the incidence plateau in 2009--11 to below 250 infections, by end of 2023 with a probability of \num{49.8}\% and by the end of 2030 with a \num{70.7}\% probability. Making the conservative assumption that denominators remain at current values, the median incidence levels for 2023 and 2030 are respectively predicted to be \num{5.36} (\num{0.398}--\num{56.3}) and \num{1.70} (\num{0.019}--110) per 10,000 at risk. These projections suggest that England will get close to a proposed elimination figure of 50 new infections per annum, which corresponds to an incidence of \num{1.1} per 10,000. Figure \ref{fig:projections} shows the distribution of the HIV incidence projections by year up to 2030, highlighting the anticipated steady decline in incidence. However, the projections become increasingly uncertain, with 40\% of the projected incidence profiles below the 1.1 level by 2030. Further investigation reveals the projections in the older age groups being the greatest contributor to this uncertainty as they are relatively more likely to have resurgent incidence in future.

Here we are benefitting from insights that can be provided by a novel age-specific CD4-staged back-calculation analysis of a rich array of HIV surveillance information using what we believe is the first example of an age-specific CD4-staged back-calculation. This approach enables estimation of age-specific HIV incidence, undiagnosed prevalence, and trends in the time to diagnosis. Although it is not possible to establish a direct causal link between these trends and the different components of combination prevention, this work enables an improved understanding of how prevention policies, particularly where they have been differentially applied by age, have shaped trends in HIV incidence. Furthermore, with the use of an appropriate smoothing on HIV incidence we have been able to derive projections of incidence to the target time for HIV elimination.

Such CD4-staged back-calculation models do have limitations. Despite consistent increases in HIV testing, a commensurate decline in time-to-diagnosis over recent periods was not found (see Fig. \ref{fig:ai.estimates}, Panel D). One explanation for this is the assumption that all CD4 counts at seroconversion are high, $>$500 cells mm$^{-3}$, ignoring the short acute infection period when CD4 counts can plummet. Those diagnosed in this period may be mis-classified by the model as long-standing infections. This mis-classification could be reduced through incorporation of information from serological tests for recent infection and dates of last negative tests when available.\cite{YanZW11} 
These additional data would also bring the further benefit of more accurate estimates of recent incidence and reduced uncertainty in the projections.

The lesson for other high-income countries from experience in England is that optimised testing and TasP have controlled country-wide the HIV epidemic in MSM. With the additional large-scale implementation of PrEP, HIV elimination, however defined, is likely to be within reach by 2030. However, targeted combination prevention measures may be needed to maintain the trajectory towards elimination in groups such as those aged 45 or over. 
Additionally, to ensure rapid and effective prevention policy adjustments, timely HIV incidence monitoring is essential to recognise and respond appropriately to changes in the current downward trend, ensuring England remains on-track for elimination.

\section*{Declaration of Interests}
No known conflicts of interest exist.

\section*{Acknowledgements}
This work was supported by the Medical Reseach Council (Unit programme number MC UU 00002/11), the UK National Institute for Health Research (NIHR) Health Protection Research Units (HPRU) on Evaluation of Interventions, both in partnership with Public Health England. The data used in the analysis are routine HIV/STI surveillance data curated by PHE and are available upon request.

\section*{Author Contributions}
FB, PB, ONG and DDA conceptualised the research, FB, PB, PK and DO were responsible for data curation, FB and PB carried out the formal analysis, FB, PB and DDA developed the methodology, FB developed software, PB, AB, VD, ONG and DDA provided supervision, FB and PB were responsible for visualisation, FB, PB and DDA wrote the original draft with all authors contributing to the reviewing and editing process.



\bibliographystyle{model2-num-names}
\bibliography{scibib.bib}







\end{document}